\documentclass[letter]{aa} 

\usepackage{graphicx}
\usepackage{color}

\newcommand{\blue}{\textcolor{black}}

\newcommand{\obsl}{\object{1E2216.0-0401}}
\newcommand{\obsr}{\object{1E2215.7-0404}}
\newcommand{\obsls}{\object{0401}}
\newcommand{\obsrs}{\object{0404}}

\usepackage{txfonts}
\begin{document} 
   \title{
   {\it Suzaku} and {\it XMM-Newton} observations of a newly-discovered \\
   early-stage cluster merger \obsl~and \obsr
   }

   \author{
   H. Akamatsu 
          \inst{1}
          \and
    L. Gu
          \inst{1}
    \and
    T. W. Shimwell \inst{2} 
    \and
    F. Mernier
  	  \inst{1,2}
    \and
    J. Mao
    	\inst{1,2}
    \and
    I. Urdampilleta 
    	\inst{1,2} 
    \and \\
    J. de Plaa
    \inst{1}
    \and 
     H.J.A. R\"ottgering \inst{2}
     \and
    J. S. Kaastra	\inst{1,2}
          }

   \institute{SRON Netherlands Institute for Space Research, Sorbonnelaan 2, 3584 CA Utrecht, The Netherlands
              \email{h.akamatsu@sron.nl}
         \and 
         Leiden Observatory, Leiden University, PO Box 9513, 2300 RA Leiden, The Netherlands
             }

   \date{}

 
\abstract{
  We present the results of {\it Suzaku} and {\it XMM-Newton} X-ray observations of the cluster pair \obsl~and \obsr. We discover an X-ray bridge between the clusters. {\it Suzaku} and {\it XMM-Newton} observations revealed that each cluster hosts gas with moderate temperature of $kT_{\obsls}=$4.8$\pm$0.1 keV and $kT_{\obsrs}=$5.8$\pm$0.2 keV, respectively. On the other hand, the bridge region shows a remarkably high temperature ({\it kT}=6.6$\pm$0.5 keV). Furthermore, at the position of the bridge, we detected an enhancement in the wavelet-decomposed soft-band (0.5--4.0 keV) {\it XMM-Newton} image at 3 sigma significance, this is  most likely due to a compression of the intracluster medium (ICM) as a consequence of the merging activity. 
This X-ray intensity and temperature enhancement are not consistent with those expected from a late phase, but are in agreement with the predictions by numerical simulations of an  early phase merger.
From the temperature jump at the location of the bridge, 
the Mach number is estimated to be ${\cal M}=1.4\pm0.1$, which corresponds to a shock propagation velocity of about 1570 km/s.  From the shock properties, we estimate that  core-passage will occur in  0.3--0.6 Gyr and that the age of the shock structure is 50--100 Myr.   
Based on the measured properties of the ICM at the bridge and estimation of timescales, we find indications for non-equilibrium ionization.  We also discover possible diffuse radio emission  located between the merging clusters. Combining the radio, X-ray, and optical image data, we speculate that the detected radio sources are most likely related to the merger event. Thus, \obsl~and \obsr~ is a new example of an early phase cluster merger with remarkable characteristics.
  }
  \authorrunning{H. Akamatsu et al.}
  \titlerunning{{\it Suzaku} and {\it XMM-Newton} X-ray observations of \obsl~and \obsr}

   \maketitle

\section{Introduction}\label{sec:intro}
According to  the framework of hierarchical structure formation,  galaxy clusters are the largest gravitationally bound objects  in the Universe. Galaxy clusters are expected to grow via merging processes i.e. from collisions with smaller sub-haloes such as sub-clusters and groups of galaxies. Such cluster merging processes release up to  $\sim10^{64}~\rm erg$ of kinetic energy into several channels such as as  heating due to merger induced shocks, turbulence and the amplification of the intergalactic magnetic field, etc.~\citep{markevitch07}. These non-thermal phenomena might (re-)acceralate  (pre-accelerated) particles up to relativistic energies, producing synchrotron radio emission~\citep[for a review, see e.g. ][]{feretti12}.  The combination of X-ray and radio observations, therefore, has the strong advantage for understanding the cluster merging process and related  phenomena. Knowledge of the cluster energy partition among different components is crucial information not only to understand its role in cluster evolution~\citep[e.g.][]{voit05} but also to construct cosmological probe~\citep[e.g.][]{allen11}.

Previously a handful of merging clusters were investigated in the X-ray band~\citep[e.g.,][]{markevitch99, markevitch02, finoguenov10, russell10, russell12,macario11, ogrean13_coma, bourdin13, akamatsu13a} and also see~\citet[][]{markevitch10}. 
However, most of these observed mergers are in a late phase (after core-crossing), because the early phase (before core-crossing) is relatively short-lived~\citep[e.g.,][]{ricker01}. 
Until now, there are  4 representative examples\footnote{We do not include A222\&223 because of their large line-of-sight distance difference$\sim$15--18 Mpc~\citep{werner08, dietrich12}.} 
of early phase cluster mergers:
A399\&A401~\citep{fujita96, fujita08}, A1750~\citep{belsole04, bulbul16}, Cygnus A~\citep{markevitch99, sarazin13} and CIZA J1358.9-4750~\citep{Kato15}. 
Although \citet{Kato15} reported a signature of a merger shock in X-ray data, the other three early phase cluster mergers do not show any clear shock structure. Furthermore, no concrete evidence of the connection between radio emission and such shocks has been reported so far. Consequently, the physics of early phase cluster mergers remains poorly understood.

In this paper, we report a new example of an early phase cluster merger based on the results of {\it Suzaku} and {\it XMM-Newton} observations of \obsl~\citep[RA = 22h18m40.5s, Dec = -03d46m48s: {\it z}=0.09:][]{cruddace02} and \obsr~\citep[RA = 22h18m17.1s, Dec = -03d50m03s: {\it z}=0.09:][]{gioia94}. 
We refer to these clusters as \obsls~and \obsrs, respectively.
Both clusters are separated by a projected distance of 640 kpc ($\sim$7.2 arcmin).
We assume the cosmological parameters $H_0 = 70$ km s$^{-1}$ Mpc$^{-1}$, $\Omega_{\rm M}=0.27$ and $\Omega_\Lambda = 0.73$. At a redshift of {\it z}=0.09, 1\arcmin~corresponds to 91.8 kpc. As our fiducial reference for the solar photospheric abundances  denoted by {\it Z}$_\odot$, we adopt \citet{lodders09}. A Galactic absorption column of $N_{\rm H}=7.3 \times 10^{20} \rm~cm^{-2}$~\citep{willingale13} was included in all fits.
Unless otherwise stated,  all our spectral analyses are done by using the Cash statistic~\citep{cash79} and the errors correspond to  68\% confidence for each parameter.

\section{Observations and data reduction}
\begin{figure}
\begin{tabular}{c}
\begin{minipage}{1\hsize}
\begin{center}
\includegraphics[width=.85\hsize]{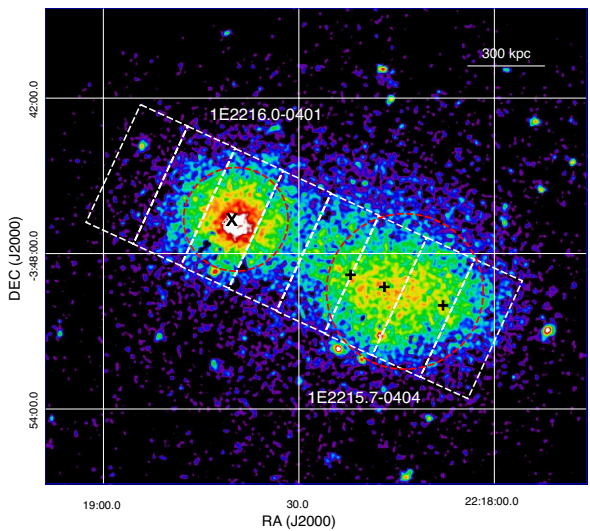}
\end{center}
\end{minipage}
\end{tabular}
\caption{\label{fig:suzaku_image}
Background-subtracted {\it XMM-Newton} image of \obsl~and \obsr~in the 0.8--8.0 keV band. 
The magenta circles and white boxes (2\arcmin $\times$5\arcmin) show the regions that we discuss in this letter. The $\times$ and + indicate the dominant galaxy(ies) of each cluster.
}
\end{figure}

{\it Suzaku}~\citep{mitsuda07} performed two observations covering both clusters (Fig.~\ref{fig:suzaku_image}).
The X-ray imaging sensor \cite[XIS:][]{koyama07} on board {\it Suzaku} consists of three front-side illuminated (FI) CCD chips (XIS0 and XIS3) and one back-side illuminated (BI) chip (XIS1).  
All observations were performed with either the normal $5\times5$ or $3\times3$ clocking mode. Data reduction was performed with HEAsoft version 6.15 and {\tt CALDB} version 20140624.
We started with the standard data screening provided by the {\it Suzaku} team and applied an event screening
 with geomagnetic cosmic-ray cut-off rigidity (COR2) $>$ 6 GV to suppress the detector background. An additional screening  was applied for the XIS1 detector to minimize the detector background. We followed the processes described in the {\it Suzaku} XIS official document\footnote{http://www.astro.isas.jaxa.jp/suzaku/analysis/xis/xis1\_ci\_6\_nxb/}. The positions of the calibration sources and the failing area in XIS0 were excluded\footnote{http://www.astro.isas.jaxa.jp/suzaku/doc/suzakumemo/suzakumemo-2010-01.pdf.}. The resultant clean exposure times are 25 and 11 ks, respectively. The observational information is shown in Tab.~\ref{tab:obslog}.

{\it XMM-Newton} was used to observe \obsl~ on 2011 June 13 for a short exposure of 15.9 ks. The SAS v13.5 and the built-in extended source analysis software (ESAS) were utilized to process and calibrate the data obtained with the {\it XMM-Newton} European Photon Imaging Camera (EPIC). Following a standard procedure, the MOS raw data were created by {\tt emchain}, and the lightcurves were extracted and screened for time variable background component by the {\tt mos-filter} task. The final net clean exposures are 10.5 ks, 13.5 ks, and 6.7 ks for the MOS1, MOS2, and pn data, respectively. The observational information is shown in Tab.~\ref{tab:obslog}.

\begin{table}
\small
\caption{\label{tab:obslog} Observational log}
\begin{center}
\begin{tabular}{lllllllllllllll}\hline
&  ID & \multicolumn{1}{l}{R.A., DEC }&  Observation & Exp\\
	&	&	(J2000) & Starting date & (ks)\\ \hline
$Suzaku$	 & 807085010 & 334.65, -3.76 & 2012-05-16 & 24.9   \\
 & 807084010 & 334.58, -3.85 & 2012-11-29 & 11.2  \\
 $XMM^a$  & 677180101& 334.57, -3.83&	2011-06-13&10.5, 13.5,6.7
 \\ \hline
\multicolumn{5}{l}{$a$: Exposure times are for EPIC MOS1, 2 and \blue{pn}} \\
\end{tabular}
\end{center}
\end{table}%

\section{Spectral analysis and result}\label{sec:spec}
\subsection{Spectral analysis approach}\label{sec:approach}
For the spectral analysis of {\it Suzaku} data, we used the SPEX~\citep{spex} software version 3.01.00.\,  Each spectrum was binned based on the ``{\tt optimal binning}'' method~\citep{kaastra16}. The best-fit parameters were obtained by minimizing the C-stat. In all the spectral analysis results presented here, the detector background (Non X-ray Background: NXB) was estimated from the night Earth data base using {\tt xisnxbgen}~\citep{tawa08}  and subtracted from the observed spectra.
Other background components such as Galactic emission and unresolved X-ray sources (Cosmic X-ray background: CXB) are modeled during the fitting procedure. For the spectral fitting, we used the energy ranges of 0.8--8.0 keV for both detectors. To estimate exactly  the responses of the X-ray telescope and XIS, we employed the Monte Carlo simulator {\tt xissimarfgen}~\citep{ishisaki07}. As an input image, we employed a flat field emission model ({\it r}=20\arcmin). For the Galactic background components, we used two thermal (the {\it cie} model in SPEX) components: one is unabsorbed and another is absorbed. In both components, we fixed the abundance and the redshift to unity and zero, respectively. Furthermore, we fixed the temperature of each component to 0.08 keV and 0.3 keV, respectively. For the fitting, we kept the normalization of thermal components free. For the CXB component, we refer to \citet{kushino02}.
In total, we used the spectral model: $cie+abs*(cie+Powerlaw+cie_{\rm ICM})$. Because of the energy range which was used for the fitting, our results are insensitive to the Galactic background components.

The {\it XMM-Newton} source spectra, NXB, and response files were prepared by the {\tt mos-spectra} task. The CXB and two Galactic foreground components were calculated in the same way for the {\it Suzaku} analysis. To complement the limited angular resolution of the {\it Suzaku} XRT~\citep[HPD$\sim$2\arcmin:][]{xrt},  we used {\it XMM-Newton} images to identify and evaluate the fluxes of point sources, which were detected by the {\tt cheese} task above a flux threshold of $5 \times 10^{-14}$ ergs cm$^{-2}$ s$^{-1}$. These point sources were excluded in all the subsequent spectral analysis.

\begin{table}
\footnotesize
\caption{{\it Suzaku} best-fit parameters for each cluster and for the bridge region\label{tab:2arcmin}}
\begin{center}
\footnotesize
\begin{tabular}{cccccccccccc} \hline
&{\it kT} &  {\it Z} & Norm & C-stat/d.o.f. \\
&(keV)	    & ({\it Z}$_\odot$) & ($10^{70}/\rm m^3/\Box\arcmin$) \\ \hline
0401&${4.75}\pm0.10  $ & $0.33\pm0.04$  & $ {31.3}\pm0.5$  &1130 / 1085\\
0404 & $  {5.68}\pm0.18  $ & $0.33\pm0.04$  & $ {18.1}\pm0.8$  & 1230 / 1229 \\
Bridge& $6.63\pm0.48$ & 0.3 (fix) & $ {14.5}\pm1.3$& 342 / 311 
\\ \hline
\end{tabular}
\end{center}
\end{table}%

\begin{figure}
\begin{tabular}{c}
\begin{minipage}{1\hsize}
\begin{center}
\includegraphics[width=1.0\hsize]{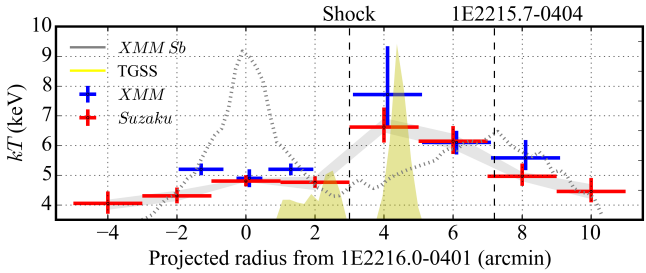}
\end{center}
\end{minipage}
\end{tabular}
\caption{\label{fig:kt}
Radial temperature profile from \obsl. Red and blue crosses show {\it Suzaku} and {\it XMM-Newton} results, respectively. The dashed line indicates the {\it XMM-Newton} 0.5--7.0 keV surface brightness profile. The gray range represent the uncertainties of {\it Suzaku} temperature estimation due to the combined 3\% variation of the NXB level and the 30\% fluctuation in the CXB.  
The yellow region shows GMRT 154 MHz radio emission.
The bins of {\it XMM-Newton} are slightly offset (0.1\arcmin) for clarity. 
}
\end{figure}

\subsection{Results}\label{sec:results}
For the first step, to understand the global properties of each cluster, we extracted spectra from circular regions centred on \obsls~and \obsrs~with {\it r}=2\arcmin~and {\it r}=3\arcmin, respectively. The best-fit values are listed in Tab.~\ref{tab:2arcmin}. Both clusters show similar values for the temperature ($kT_{\obsls}=4.8\pm0.1~{\rm keV},~ kT_{\obsrs}=\rm 5.7\pm0.2~{\rm keV}$) and the abundance ($Z_{\obsls}=0.33\pm0.04 ~{\it Z}_\odot, Z_{\obsrs}=0.33\pm0.04~{\it Z}_\odot$). The redshifts derived for the clusters are $z_{\obsls}=0.093\pm0.001$ and $z_{\obsrs}=0.091\pm0.001$.
The velocity difference derived from the redshifts $\Delta$v$\sim 600~\rm km/s$ is smaller than the sound speed (galaxy velocity dispersion) of each cluster ($c_s\sim1100~\rm km/s$). Thus the two clusters are at a similar line-of-sight distance.

In order to investigate the temperature structure in the direction of the merging axis (\obsls~to \obsrs), we extracted spectra in 8 boxes (2\arcmin$\times$5\arcmin: 182 kpc$\times$460 kpc), as shown in Fig.~\ref{fig:suzaku_image}. We followed the same procedure described above but fixed the abundance to 0.3{\it Z}$_\odot$~{\citep[e.g., ][]{fujita08}.}\footnote{Considering the difference between the two solar abundance tables, the Fe abundance in this latter is 1.5 times higher than that of \cite{fujita08}.}  ~
To investigate the influence of uncertainties in the background components~\blue{in the {\it Suzaku} data}, we considered 30\% and 3\% fluctuations of the intensity of the CXB and the NXB components~\citep{tawa08}. Because we did not use the low-energy band, the effect of the Galactic background is limited. Therefore, we did not consider this systematic error. 

\begin{figure*}
\begin{tabular}{c}
\begin{minipage}{.5\hsize}
\begin{center}
\includegraphics[width=.9\hsize]{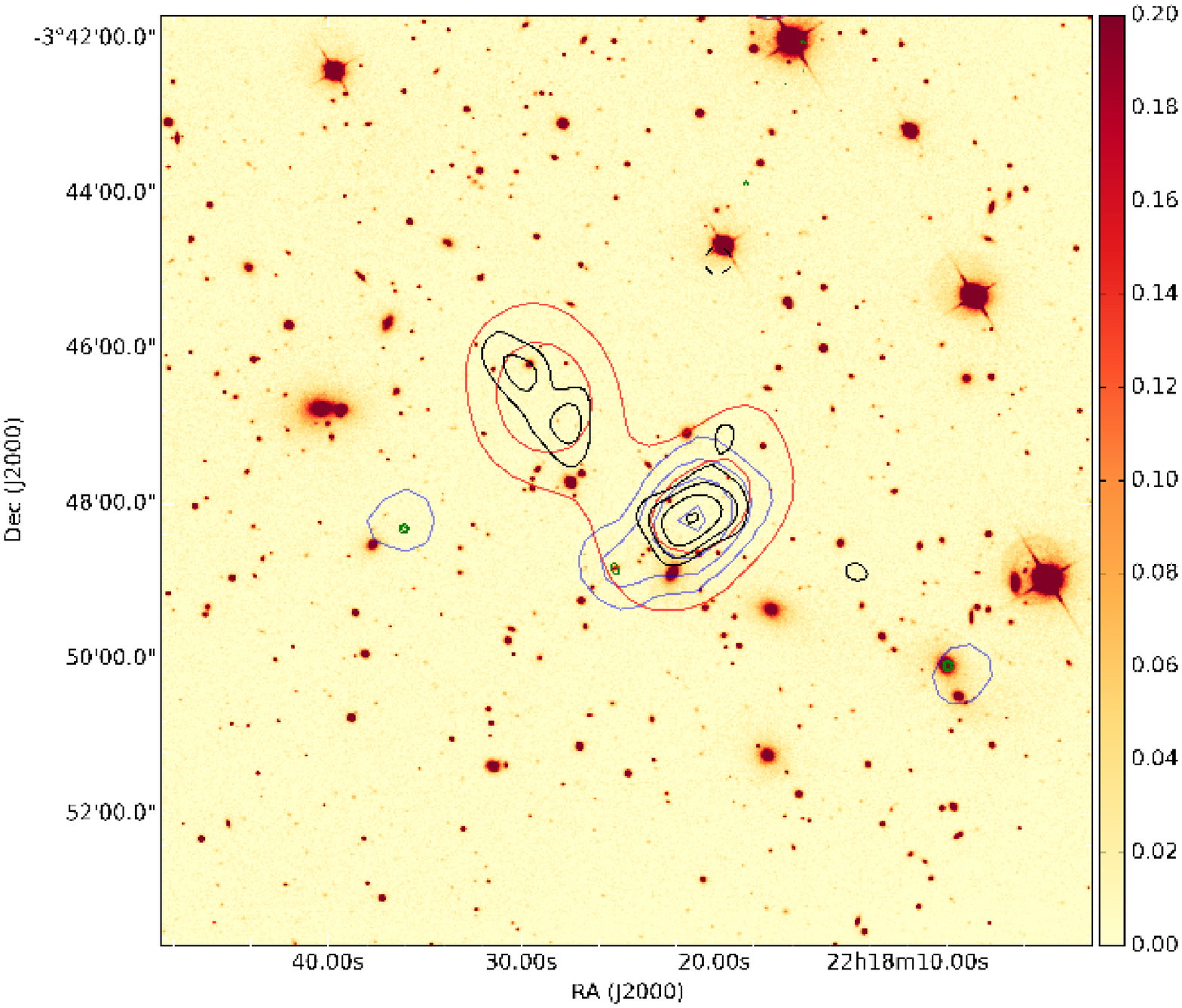}
\end{center}
\end{minipage}
\begin{minipage}{.5\hsize}
\begin{center}
\includegraphics[width=.9\hsize]{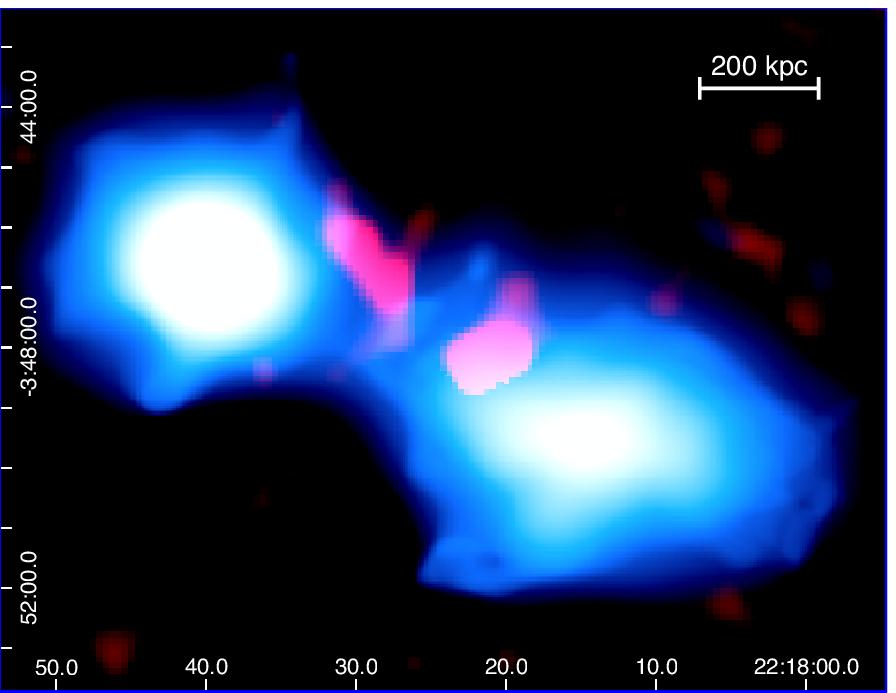}
\end{center}
\end{minipage}
\end{tabular}
\caption{\label{fig:wavlet}
Left: the radio emission from VLSSr, TGSS, FIRST, and NVSS in red, black, green and blue respectively. The contours are at $(1,2,4,...)\times 4 \times \sigma$ levels where $\sigma$ is 90~mJy/beam, 6.0~mJy/beam, 0.24~mJy/beam and 0.6~mJy/beam for the VLSSr, TGSS, FIRST and NVSS respectively. The radio emission is overlaid on an SDSS r-band image of the region.
Right:  wavelet-decomposed 0.5--4.0 keV {\it XMM-Newton} image in blue.
We show sources with $> 3\sigma$ significance and scales above$\sim$100 kpc.
The filamentary structure between two halos is visible in the image.
The red shows the 154MHz radio intensity from the TGSS alternative data release~\citep[][]{tgss}.
Furthermore, there is a structure in the X-ray image at the bridge, which might be the shock compressed ICM region.
}
\end{figure*}

The resultant temperature profile is shown in Fig.~\ref{fig:kt}, where red and blue crosses represent the best fit value of {\it Suzaku} and {\it XMM-Newton}, respectively.  In general, the results of {\it Suzaku} and {\it XMM-Newton} are consistent with each other within the statistical uncertainties ($68~\%$ significance). The basic feature of the profile is insensitive to the systematic error of the background components (gray shallow region). Between both clusters ({\it r}=3.0\arcmin--5.0\arcmin), the ICM shows a clear enhancement from $kT_{r=1\arcmin-3\arcmin}=4.7$ keV in the cluster centres to $kT_{Bridge, Suzaku}=6.6$ keV in the region between the clusters. We also found \blue{an enhancement of the surface brightness} in the 0.5--4.0 keV {\it XMM-Newton} image (Fig.~\ref{fig:kt} and \ref{fig:wavlet}).

To detect enhanced ICM structure in the connecting region, we apply a wavelet decomposition based on the $\rm \grave{a}$ Trous transform \cite[e.g.,][]{gu09} on the vignetting-corrected {\it XMM-Newton} image. By masking out coefficients within 3$\sigma$ fluctuations of the background region, and reconstructing the features with scales above $\sim$ 100 kpc, we obtained a significant coefficient map as shown in Fig.~\ref{fig:wavlet} right.
The observed features, such as the shock heated and compressed region between both clusters, are in good agreement with predictions by numerical simulations ~\citep[][]{takizawa99, akahori10}.

The structure located between the clusters and centered on the high temperature region, suggests the presence of a pressure discontinuity, that is, the presence of a shock front.

\section{Discussion}\label{sec:discussion}
The observation of a clear shock structure in an early phase cluster merger is not so common so far.
In order to understand the nature of early phase cluster mergers, additional examples are of importance. 

The observational features revealed by {\it Suzaku} and {\it XMM-Newton} can not be explained by the late phase of a merger. Furthermore,  based on available optical information, we found the presence of some peculiar galaxies in~\obsls~and \obsrs. The positions of these galaxies are roughly consistent with the X-ray peak of each cluster ($\times$ and + in Fig.~\ref{fig:suzaku_image}). This indicates that the ICM of each cluster has not been perturbed for a long time by the merger.  Most likely the clusters are before core-crossing (early merger phase)  but when they are close enough to interact and form a shock front.

The {\it Suzaku} and {\it XMM-Newton} temperature profiles show a clear jump at the bridge region, indicating the presence of a shock front. Here we evaluate the properties of this shock structure based only on the {\it Suzaku} measurements because of they have a  lower particle background and longer observation time.
The Mach number ${\cal M}$ can be estimated by applying the Rankine- Hugoniot jump condition~\citep{landau59_fluid}, 
$\displaystyle{
\frac{T_2}{T_1} = \frac{5{\cal M}^4+14{\cal M}^2-3}{16{\cal M}^2}
}
$
assuming the ratio of specific heats as $\gamma=5/3$ and 1,2 denote  pre-shock and post-shock, respectively. Substituting the pre- and post-shock temperature $T_1=4.7\pm0.2$ keV and $T_2=6.6\pm0.5$ keV into the above equation gives a Mach number ${\cal M}_{Suzaku}=1.4\pm0.1$. 
Following equation (2) in \citet{markevitch99}, the shock compression {\it C} can be estimated from the measured temperature ratio.  Substituting the pre- and post-shock temperature ratio into the equation gives a shock compression  $C\sim1.56$. The heating due to adiabatic compression is expected to be about $T_2=T_1\cdot C^{\gamma-1}\sim6.3$ keV, in good agreement with the post-shock temperature. Thus, the dominant part of the heating is carried out by adiabatic compression. The estimated low Mach number and heating mechanism are consistent with the properties of a shock structure that cosmological simulations predict to occur in an early-phase cluster merger~\citep[e.g., ][]{miniati00, ricker01, ryu03}.

Combining the sound speed at the pre-shock region ($c_s\sim1150~\rm km~s^{-1}$) with the Mach number, the shock propagation velocity is estimated to be about 1570 km/s. With the assumption that the clusters move with the same velocity, we estimate that ``the core passing" will occur in  0.3--0.6 Gyr. Under the same assumption, the age of the shock is also estimated to be 50--100 Myr.

Due to the low density environment at cluster outskirts, it is commonly believed that non-equilibrium ionization and electron-ion temperature structure remain just after the shock because it has not had enough time to equilibrate~\citep{takizawa98, rudd09, akahori10}. 
The equilibration time scales of non-equilibrium ionization $t_{\rm CIE}$ and electron-ion two temperature structure $t_{\rm ie}$~\citep{ spitzer56, masai84, takizawa98b} can be described as
$\displaystyle{t_{\rm CIE}}\sim3\times10^{9}(\frac{n_{\rm e}}{10^{-4}~\rm cm^{-3}})^{-1} \rm yr$, 
$\displaystyle{t_{\rm ie}\sim2\times10^{9}(\frac{n_{\rm e}}{10^{-4}~\rm cm^{-3}})^{-1}(\frac{T}{10^8\rm K})^{\frac{3}{2}}}\rm ~yr $, respectively. 
Here, we estimate the electron density of the bridge region from the {\it cie} normalization ($norm=n_{\rm e}n_{\rm X}V$).
Assuming a 2 Mpc line of sight depth and the normalization value of the unit area (Tab.\ref{tab:2arcmin}), the electron density at the shock heated region is estimated to be 
$\displaystyle{n_{\rm e}=6.0\times10^{-4}~\rm ~({\it l}/2 Mpc)^{-0.5}cm^3}$, where {\it l} is the line-of-sight depth. We note that {\it l} is quite unconstrained, hence $n_e$ is rather uncertain.
Combining with the assumption of $l=2$ Mpc and the measured shock temperature, the  equilibration timescales are estimated to be $t_{\rm CIE}\sim$0.5 \rm~Gyr and $t_{\rm ie}\sim$0.2 Gyr. These timescales are longer than the estimated age of the shock structure estimated from the shock propagation velocity, indicating a part of the shock heated region would not have reached equilibrium yet.

To investigate the possibility of a non-equilibrium ionization state, we evaluated the {\it rt} parameter in the {\it cie} model, which represents  the ratio of ionization balance temperature to electron temperature. The resultant value  (${\it rt}=0.45_{-0.10}^{+0.38}$) suggests that the ICM at the shock  region has not yet reached ionization equilibrium, which will lead to an underestimation of the post-shock temperature.  Furthermore we note that {\it Suzaku}'s large PSF might lead to a similar effect due to the effects of smearing.  Therefore, the estimated Mach number is a lower limit. We stress that it is a challenge to make a firm conclusion about non-equilibrium ionization states with the currently available  X-ray data. Deep X-ray observations are strongly desired. Future high resolution X-ray spectrometers such as the {\it Athena} X-IFU instrument \citep[][]{xifu14} will enable us not only to investigate these non-equilibrium states but also detailed spectroscopic diagnostics~\citep[e.g., ][]{kaastra09}.

Diffuse radio emission is observed to be associated with the ICM of approximately 100 galaxy clusters and is generally thought to be formed as a consequence of shocks and turbulence in massive clusters (see e.g.  \citealt{feretti12} and \citealt{brunetti14} for recent reviews). To search for such diffuse radio emission from the ICM of this merging cluster system and to characterise it as a function of frequency and resolution we have used the following wide-area radio surveys: the 74\,MHz VLA Low-Frequency Sky Survey (VLSSr; \citealt{lane12}), the 150\,MHz TIFR GMRT Sky Survey alternative data release (TGSS; \citealt{tgss}), the 1.4\,GHz NRAO/VLA Sky Survey (NVSS; \citealt{condon98}), and the 1.4\,GHz Faint Radio Images of the Sky at Twenty-Centimeters (FIRST; \citealt{becker95}). We find that in the vicinity of 0401 and 0404 there are two main regions of diffuse radio emission which are approximately coincident with the X-ray observed bridge between the clusters, these are object A at 22:18:22 -03:48:08 and object B at 22:18:29 -03:46:42.  From the radio survey images we estimate that for object A the 150\,MHz integrated brightness is $570\pm$60\,mJy and the 74\,MHz to 1.4\,GHz spectral index is -1.2$\pm$0.1, whereas for object B the integrated 150\,MHz brightness is 350$\pm$50\,mJy and the 74\,MHz to 150\,MHz spectral index is -2.0$\pm$0.3 (object B is not detected in the 1.4GHz images). Steep spectral indices together with the lack of an obvious optical counterpart or compact radio core (see Fig.\ref{fig:wavlet} left) are properties that are expected for radio emission that is associated with or influenced by shocks and turbulence in the ICM. Furthermore, such a classification is supported by the clear merging nature of the system and the close proximity of the radio emission and the X-ray detected shock. However,  before any definitive classification of this radio emission is made it is important that more sensitive and higher resolution radio observations are conducted to thoroughly assess the morphological, spectral and polarisation properties of the emission.

In this paper we \blue{showed} that \obsl~and \obsr~are a new example of an early phase cluster merger with a clear signature of a merger induced shock front and possibility of  diffuse radio emission which also may relate to the merging activity.  Further investigation with high spatial and spectral resolution X-ray observatories ({\it XMM-Newton}, {\it Chandra} and also {\it Athena}), sensitive radio observations (GMRT, JVLA and LOFAR), weak lensing mass maps, and sophisticated numerical simulations will provide us with deeper insight of the cluster merger dynamics and the origin of the diffuse radio emission.

\section*{Acknowledgments}
The authors wish to thank the referee for constructive comments that improved the manuscript.
The authors thank the {\it Suzaku} and {\it XMM-Newton} team members for their support on the projects.
H.A acknowledges the support of NWO via a Veni grant.
SRON is supported financially by NWO, the Netherlands Organization for Scientific Research. 

\bibliographystyle{aa}
\bibliography{ref}

\begin{thebibliography}{54}
\expandafter\ifx\csname natexlab\endcsname\relax\def\natexlab#1{#1}\fi

\bibitem[{{Akahori} \& {Yoshikawa}(2010)}]{akahori10}
{Akahori}, T. \& {Yoshikawa}, K. 2010, \pasj, 62, 335

\bibitem[{{Akamatsu} \& {Kawahara}(2013)}]{akamatsu13a}
{Akamatsu}, H. \& {Kawahara}, H. 2013, \pasj, 65, 16

\bibitem[{{Allen} {et~al.}(2011){Allen}, {Evrard}, \& {Mantz}}]{allen11}
{Allen}, S.~W., {Evrard}, A.~E., \& {Mantz}, A.~B. 2011, \araa, 49, 409

\bibitem[{{Becker} {et~al.}(1995){Becker}, {White}, \& {Helfand}}]{becker95}
{Becker}, R.~H., {White}, R.~L., \& {Helfand}, D.~J. 1995, \apj, 450, 559

\bibitem[{{Belsole} {et~al.}(2004){Belsole}, {Pratt}, {Sauvageot}, \&
  {Bourdin}}]{belsole04}
{Belsole}, E., {Pratt}, G.~W., {Sauvageot}, J.-L., \& {Bourdin}, H. 2004, \aap,
  415, 821

\bibitem[{{Bourdin} {et~al.}(2013){Bourdin}, {Mazzotta}, {Markevitch},
  {Giacintucci}, \& {Brunetti}}]{bourdin13}
{Bourdin}, H., {Mazzotta}, P., {Markevitch}, M., {Giacintucci}, S., \&
  {Brunetti}, G. 2013, \apj, 764, 82

\bibitem[{{Brunetti} \& {Jones}(2014)}]{brunetti14}
{Brunetti}, G. \& {Jones}, T.~W. 2014, International Journal of Modern Physics
  D, 23, 30007

\bibitem[{{Bulbul} {et~al.}(2016){Bulbul}, {Randall}, {Bayliss}, {Miller},
  {Andrade-Santos}, {Johnson}, {Bautz}, {Blanton}, {Forman}, {Jones},
  {Paterno-Mahler}, {Murray}, {Sarazin}, {Smith}, \& {Ezer}}]{bulbul16}
{Bulbul}, E., {Randall}, S.~W., {Bayliss}, M., {et~al.} 2016, \apj, 818, 131

\bibitem[{{Cash}(1979)}]{cash79}
{Cash}, W. 1979, \apj, 228, 939

\bibitem[{{Condon} {et~al.}(1998){Condon}, {Cotton}, {Greisen}, {Yin},
  {Perley}, {Taylor}, \& {Broderick}}]{condon98}
{Condon}, J.~J., {Cotton}, W.~D., {Greisen}, E.~W., {et~al.} 1998, \aj, 115,
  1693

\bibitem[{{Cruddace} {et~al.}(2002){Cruddace}, {Voges}, {B{\"o}hringer},
  {Collins}, {Romer}, {MacGillivray}, {Yentis}, {Schuecker}, {Ebeling}, \& {De
  Grandi}}]{cruddace02}
{Cruddace}, R., {Voges}, W., {B{\"o}hringer}, H., {et~al.} 2002, \apjs, 140,
  239

\bibitem[{{Dietrich} {et~al.}(2012){Dietrich}, {Werner}, {Clowe}, {Finoguenov},
  {Kitching}, {Miller}, \& {Simionescu}}]{dietrich12}
{Dietrich}, J.~P., {Werner}, N., {Clowe}, D., {et~al.} 2012, \nat, 487, 202

\bibitem[{{Feretti} {et~al.}(2012){Feretti}, {Giovannini}, {Govoni}, \&
  {Murgia}}]{feretti12}
{Feretti}, L., {Giovannini}, G., {Govoni}, F., \& {Murgia}, M. 2012, \aapr, 20,
  54

\bibitem[{{Finoguenov} {et~al.}(2010){Finoguenov}, {Sarazin}, {Nakazawa},
  {Wik}, \& {Clarke}}]{finoguenov10}
{Finoguenov}, A., {Sarazin}, C.~L., {Nakazawa}, K., {Wik}, D.~R., \& {Clarke},
  T.~E. 2010, \apj, 715, 1143

\bibitem[{{Fujita} {et~al.}(1996){Fujita}, {Koyama}, {Tsuru}, \&
  {Matsumoto}}]{fujita96}
{Fujita}, Y., {Koyama}, K., {Tsuru}, T., \& {Matsumoto}, H. 1996, \pasj, 48,
  191

\bibitem[{{Fujita} {et~al.}(2008){Fujita}, {Tawa}, {Hayashida}, {Takizawa},
  {Matsumoto}, {Okabe}, \& {Reiprich}}]{fujita08}
{Fujita}, Y., {Tawa}, N., {Hayashida}, K., {et~al.} 2008, \pasj, 60, 343

\bibitem[{{Gioia} \& {Luppino}(1994)}]{gioia94}
{Gioia}, I.~M. \& {Luppino}, G.~A. 1994, \apjs, 94, 583

\bibitem[{{Gu} {et~al.}(2009){Gu}, {Xu}, {Gu}, {Wang}, {Zhang}, {Wang}, {Qin},
  {Cui}, \& {Wu}}]{gu09}
{Gu}, L., {Xu}, H., {Gu}, J., {et~al.} 2009, \apj, 700, 1161

\bibitem[{{Intema} {et~al.}(2016){Intema}, {Jagannathan}, {Mooley}, \&
  {Frail}}]{tgss}
{Intema}, H.~T., {Jagannathan}, P., {Mooley}, K.~P., \& {Frail}, D.~A. 2016,
  ArXiv e-prints

\bibitem[{{Ishisaki} {et~al.}(2007){Ishisaki}, {Maeda}, {Fujimoto}, {Ozaki},
  {Ebisawa}, {Takahashi}, {Ueda}, {Ogasaka}, {Ptak}, {Mukai}, {Hamaguchi},
  {Hirayama}, {Kotani}, {Kubo}, {Shibata}, {Ebara}, {Furuzawa}, {Iizuka},
  {Inoue}, {Mori}, {Okada}, {Yokoyama}, {Matsumoto}, {Nakajima}, {Yamaguchi},
  {Anabuki}, {Tawa}, {Nagai}, {Katsuda}, {Hayashida}, {Bamba}, {Miller},
  {Sato}, \& {Yamasaki}}]{ishisaki07}
{Ishisaki}, Y., {Maeda}, Y., {Fujimoto}, R., {et~al.} 2007, \pasj, 59, 113

\bibitem[{{Kaastra} \& {Bleeker}(2016)}]{kaastra16}
{Kaastra}, J.~S. \& {Bleeker}, J.~A.~M. 2016, \aap, 587, A151

\bibitem[{{Kaastra} {et~al.}(2009){Kaastra}, {Bykov}, \& {Werner}}]{kaastra09}
{Kaastra}, J.~S., {Bykov}, A.~M., \& {Werner}, N. 2009, \aap, 503, 373

\bibitem[{{Kaastra} {et~al.}(1996){Kaastra}, {Mewe}, \&
  {Nieuwenhuijzen}}]{spex}
{Kaastra}, J.~S., {Mewe}, R., \& {Nieuwenhuijzen}, H. 1996, in UV and X-ray
  Spectroscopy of Astrophysical and Laboratory Plasmas, ed. K.~{Yamashita} \&
  T.~{Watanabe}, 411--414

\bibitem[{{Kato} {et~al.}(2015){Kato}, {Nakazawa}, {Gu}, {Akahori}, {Takizawa},
  {Fujita}, \& {Makishima}}]{Kato15}
{Kato}, Y., {Nakazawa}, K., {Gu}, L., {et~al.} 2015, \pasj, 67, 71

\bibitem[{{Koyama} {et~al.}(2007){Koyama}, {Tsunemi}, {Dotani}, {Bautz},
  {Hayashida}, {Tsuru}, {Matsumoto}, {Ogawara}, {Ricker}, {Doty}, {Kissel},
  {Foster}, {Nakajima}, {Yamaguchi}, {Mori}, {Sakano}, {Hamaguchi},
  {Nishiuchi}, {Miyata}, {Torii}, {Namiki}, {Katsuda}, {Matsuura}, {Miyauchi},
  {Anabuki}, {Tawa}, {Ozaki}, {Murakami}, {Maeda}, {Ichikawa}, {Prigozhin},
  {Boughan}, {Lamarr}, {Miller}, {Burke}, {Gregory}, {Pillsbury}, {Bamba},
  {Hiraga}, {Senda}, {Katayama}, {Kitamoto}, {Tsujimoto}, {Kohmura}, {Tsuboi},
  \& {Awaki}}]{koyama07}
{Koyama}, K., {Tsunemi}, H., {Dotani}, T., {et~al.} 2007, \pasj, 59, 23

\bibitem[{{Kushino} {et~al.}(2002){Kushino}, {Ishisaki}, {Morita}, {Yamasaki},
  {Ishida}, {Ohashi}, \& {Ueda}}]{kushino02}
{Kushino}, A., {Ishisaki}, Y., {Morita}, U., {et~al.} 2002, \pasj, 54, 327

\bibitem[{{Landau} \& {Lifshitz}(1959)}]{landau59_fluid}
{Landau}, L.~D. \& {Lifshitz}, E.~M. 1959, {Fluid mechanics}

\bibitem[{{Lane} {et~al.}(2012){Lane}, {Cotton}, {Helmboldt}, \&
  {Kassim}}]{lane12}
{Lane}, W.~M., {Cotton}, W.~D., {Helmboldt}, J.~F., \& {Kassim}, N.~E. 2012,
  Radio Science, 47, RS0K04

\bibitem[{{Lodders} {et~al.}(2009){Lodders}, {Palme}, \& {Gail}}]{lodders09}
{Lodders}, K., {Palme}, H., \& {Gail}, H.-P. 2009, Landolt B{\"o}rnstein

\bibitem[{{Macario} {et~al.}(2011){Macario}, {Markevitch}, {Giacintucci},
  {Brunetti}, {Venturi}, \& {Murray}}]{macario11}
{Macario}, G., {Markevitch}, M., {Giacintucci}, S., {et~al.} 2011, \apj, 728,
  82

\bibitem[{{Markevitch}(2010)}]{markevitch10}
{Markevitch}, M. 2010, ArXiv e-prints

\bibitem[{{Markevitch} {et~al.}(2002){Markevitch}, {Gonzalez}, {David},
  {Vikhlinin}, {Murray}, {Forman}, {Jones}, \& {Tucker}}]{markevitch02}
{Markevitch}, M., {Gonzalez}, A.~H., {David}, L., {et~al.} 2002, \apjl, 567,
  L27

\bibitem[{{Markevitch} {et~al.}(1999){Markevitch}, {Sarazin}, \&
  {Vikhlinin}}]{markevitch99}
{Markevitch}, M., {Sarazin}, C.~L., \& {Vikhlinin}, A. 1999, \apj, 521, 526

\bibitem[{{Markevitch} \& {Vikhlinin}(2007)}]{markevitch07}
{Markevitch}, M. \& {Vikhlinin}, A. 2007, \physrep, 443, 1

\bibitem[{{Masai}(1984)}]{masai84}
{Masai}, K. 1984, \apss, 98, 367

\bibitem[{{Miniati} {et~al.}(2000){Miniati}, {Ryu}, {Kang}, {Jones}, {Cen}, \&
  {Ostriker}}]{miniati00}
{Miniati}, F., {Ryu}, D., {Kang}, H., {et~al.} 2000, \apj, 542, 608

\bibitem[{{Mitsuda} {et~al.}(2007){Mitsuda}, {Bautz}, {Inoue}, {Kelley},
  {Koyama}, {Kunieda}, {Makishima}, {Ogawara}, {Petre}, {Takahashi}, {Tsunemi},
  {White}, {Anabuki}, {Angelini}, {Arnaud}, {Awaki}, {Bamba}, {Boyce}, {Brown},
  {Chan}, {Cottam}, {Dotani}, {Doty}, {Ebisawa}, {Ezoe}, {Fabian}, {Figueroa},
  {Fujimoto}, {Fukazawa}, {Furusho}, {Furuzawa}, {Gendreau}, {Griffiths},
  {Haba}, {Hamaguchi}, {Harrus}, {Hasinger}, {Hatsukade}, {Hayashida}, {Henry},
  {Hiraga}, {Holt}, {Hornschemeier}, {Hughes}, {Hwang}, {Ishida}, {Ishisaki},
  {Isobe}, {Itoh}, {Iyomoto}, {Kahn}, {Kamae}, {Katagiri}, {Kataoka},
  {Katayama}, {Kawai}, {Kilbourne}, {Kinugasa}, {Kissel}, {Kitamoto}, {Kohama},
  {Kohmura}, {Kokubun}, {Kotani}, {Kotoku}, {Kubota}, {Madejski}, {Maeda},
  {Makino}, {Markowitz}, {Matsumoto}, {Matsumoto}, {Matsuoka}, {Matsushita},
  {McCammon}, {Mihara}, {Misaki}, {Miyata}, {Mizuno}, {Mori}, {Mori}, {Morii},
  {Moseley}, {Mukai}, {Murakami}, {Murakami}, {Mushotzky}, {Nagase}, {Namiki},
  {Negoro}, {Nakazawa}, {Nousek}, {Okajima}, {Ogasaka}, {Ohashi}, {Oshima},
  {Ota}, {Ozaki}, {Ozawa}, {Parmar}, {Pence}, {Porter}, {Reeves}, {Ricker},
  {Sakurai}, {Sanders}, {Senda}, {Serlemitsos}, {Shibata}, {Soong}, {Smith},
  {Suzuki}, {Szymkowiak}, {Takahashi}, {Tamagawa}, {Tamura}, {Tamura},
  {Tanaka}, {Tashiro}, {Tawara}, {Terada}, {Terashima}, {Tomida}, {Torii},
  {Tsuboi}, {Tsujimoto}, {Tsuru}, {Turner}, {Ueda}, {Ueno}, {Ueno}, {Uno},
  {Urata}, {Watanabe}, {Yamamoto}, {Yamaoka}, {Yamasaki}, {Yamashita},
  {Yamauchi}, {Yamauchi}, {Yaqoob}, {Yonetoku}, \& {Yoshida}}]{mitsuda07}
{Mitsuda}, K., {Bautz}, M., {Inoue}, H., {et~al.} 2007, \pasj, 59, 1

\bibitem[{{Ogrean} \& {Br{\"u}ggen}(2013)}]{ogrean13_coma}
{Ogrean}, G.~A. \& {Br{\"u}ggen}, M. 2013, \mnras, 433, 1701

\bibitem[{{Ravera} {et~al.}(2014){Ravera}, {Barret}, {den Herder}, {Piro},
  {Cl{\'e}dassou}, {Pointecouteau}, {Peille}, {Pajot}, {Arnaud}, {Pigot},
  {Duband}, {Cara}, {den Hartog}, {Gottardi}, {Akamatsu}, {van der Kuur}, {van
  Weers}, {de Plaa}, {Macculi}, {Lotti}, {Torrioli}, {Gatti}, {Valenziano},
  {Barbera}, {Barcons}, {Ceballos}, {F{\`a}brega}, {Mas-Hesse}, {Page},
  {Guttridge}, {Willingale}, {Paltani}, {Genolet}, {Bozzo}, {Rauw}, {Renotte},
  {Wilms}, \& {Schmid}}]{xifu14}
{Ravera}, L., {Barret}, D., {den Herder}, J.~W., {et~al.} 2014, in \procspie,
  Vol. 9144, Space Telescopes and Instrumentation 2014: Ultraviolet to Gamma
  Ray, 91442L

\bibitem[{{Ricker} \& {Sarazin}(2001)}]{ricker01}
{Ricker}, P.~M. \& {Sarazin}, C.~L. 2001, \apj, 561, 621

\bibitem[{{Rudd} \& {Nagai}(2009)}]{rudd09}
{Rudd}, D.~H. \& {Nagai}, D. 2009, \apjl, 701, L16

\bibitem[{{Russell} {et~al.}(2012){Russell}, {McNamara}, {Sanders}, {Fabian},
  {Nulsen}, {Canning}, {Baum}, {Donahue}, {Edge}, {King}, \&
  {O'Dea}}]{russell12}
{Russell}, H.~R., {McNamara}, B.~R., {Sanders}, J.~S., {et~al.} 2012, \mnras,
  423, 236

\bibitem[{{Russell} {et~al.}(2010){Russell}, {Sanders}, {Fabian}, {Baum},
  {Donahue}, {Edge}, {McNamara}, \& {O'Dea}}]{russell10}
{Russell}, H.~R., {Sanders}, J.~S., {Fabian}, A.~C., {et~al.} 2010, \mnras,
  406, 1721

\bibitem[{{Ryu} {et~al.}(2003){Ryu}, {Kang}, {Hallman}, \& {Jones}}]{ryu03}
{Ryu}, D., {Kang}, H., {Hallman}, E., \& {Jones}, T.~W. 2003, \apj, 593, 599

\bibitem[{{Sarazin} {et~al.}(2013){Sarazin}, {Finoguenov}, \&
  {Wik}}]{sarazin13}
{Sarazin}, C.~L., {Finoguenov}, A., \& {Wik}, D.~R. 2013, Astronomische
  Nachrichten, 334, 346

\bibitem[{{Serlemitsos} {et~al.}(2007){Serlemitsos}, {Soong}, {Chan},
  {Okajima}, {Lehan}, {Maeda}, {Itoh}, {Mori}, {Iizuka}, {Itoh}, {Inoue},
  {Okada}, {Yokoyama}, {Itoh}, {Ebara}, {Nakamura}, {Suzuki}, {Ishida},
  {Hayakawa}, {Inoue}, {Okuma}, {Kubota}, {Suzuki}, {Osawa}, {Yamashita},
  {Kunieda}, {Tawara}, {Ogasaka}, {Furuzawa}, {Tamura}, {Shibata}, {Haba},
  {Naitou}, \& {Misaki}}]{xrt}
{Serlemitsos}, P.~J., {Soong}, Y., {Chan}, K.-W., {et~al.} 2007, \pasj, 59, 9

\bibitem[{{Spitzer}(1956)}]{spitzer56}
{Spitzer}, L. 1956, {Physics of Fully Ionized Gases}

\bibitem[{{Takizawa}(1998)}]{takizawa98b}
{Takizawa}, M. 1998, \apj, 509, 579

\bibitem[{{Takizawa}(1999)}]{takizawa99}
{Takizawa}, M. 1999, \apj, 520, 514

\bibitem[{{Takizawa} \& {Mineshige}(1998)}]{takizawa98}
{Takizawa}, M. \& {Mineshige}, S. 1998, \apj, 499, 82

\bibitem[{{Tawa} {et~al.}(2008){Tawa}, {Hayashida}, {Nagai}, {Nakamoto},
  {Tsunemi}, {Yamaguchi}, {Ishisaki}, {Miller}, {Mizuno}, {Dotani}, {Ozaki}, \&
  {Katayama}}]{tawa08}
{Tawa}, N., {Hayashida}, K., {Nagai}, M., {et~al.} 2008, \pasj, 60, 11

\bibitem[{{Voit}(2005)}]{voit05}
{Voit}, G.~M. 2005, Reviews of Modern Physics, 77, 207

\bibitem[{{Werner} {et~al.}(2008){Werner}, {Finoguenov}, {Kaastra},
  {Simionescu}, {Dietrich}, {Vink}, \& {B{\"o}hringer}}]{werner08}
{Werner}, N., {Finoguenov}, A., {Kaastra}, J.~S., {et~al.} 2008, \aap, 482, L29

\bibitem[{{Willingale} {et~al.}(2013){Willingale}, {Starling}, {Beardmore},
  {Tanvir}, \& {O'Brien}}]{willingale13}
{Willingale}, R., {Starling}, R.~L.~C., {Beardmore}, A.~P., {Tanvir}, N.~R., \&
  {O'Brien}, P.~T. 2013, \mnras, 431, 394

\end{thebibliography}
\appendix
\label{lastpage}

\end{document}